# How Good is Your Data? Investigating the Quality of Data Generated During Security Incident Response Investigations

George Grispos
School of Interdisciplinary Informatics
College of Information Science and Technology
University of Nebraska at Omaha
ggrispos@unomaha.edu

William Bradley Glisson
Cyber Forensics Intelligence Center, Department of Computer Science
Sam Houston State University
glisson@shsu.edu

Tim Storer
School of Computing Science
College of Science and Engineering
University of Glasgow
timothy.storer@glasgow.ac.uk

## Abstract

*An increasing number of cybersecurity incidents prompts organizations to explore alternative security solutions, such as threat intelligence programs. For such programs to succeed, data needs to be collected, validated, and recorded in relevant datastores. One potential source supplying these datastores is an organization's security incident response team. However, researchers have argued that these teams focus more on eradication and recovery and less on providing feedback to enhance organizational security. This prompts the idea that data collected during security incident investigations may be of insufficient quality for threat intelligence analysis.*

*While previous discussions focus on data quality issues from threat intelligence sharing perspectives, minimal research examines the data generated during incident response investigations. This paper presents the results of a case study identifying data quality challenges in a Fortune 500 organization's incident response team. Furthermore, the paper provides the foundation for future research regarding data quality concerns in security incident response.*

## 1. Introduction

Cybersecurity incidents continue to plague organizations around the world [1-4]. According to the 2017 SANS Institute Incident Response Survey [1], 87% of respondents detected at least one security incident in the past year. Similarly, the United Kingdom (UK) Government's 2017 Cyber Security Breaches Survey [2] reported that 46% of all UK business identified at least one security breach or attack in the previous twelve months. These security attacks and breaches come at a tremendous financial cost. For example, the Ponemon Institute estimates that economic losses attributed to security incidents and breaches cost United States-based organizations an average of $7.35 million in 2016 [3].

The increasing number of security attacks and data breaches fuels regulatory legislation and directives. These directives mandate that organizations implement mechanisms to protect, recover from and investigate attacks that become security incidents. For example, Article 32.1(c) within the European Union's General Data Protection Regulation (GDPR) specifies that organizations must have "the ability to restore the availability and access to personal data in a timely manner in the event of a physical or technical incident" [5]. Similarly, in the United States, forty-eight states and the District of Columbia have enacted legislation requiring organizations to notify individuals when security breaches involve personally identifiable information [6].

In an effort to assist organizations with legal and regulatory obligations, several institutions, such as the National Institute of Standards and Technology (NIST) [7], the European Network and Information Security Agency (ENISA) [8] and the International Organization for Standardization (ISO) [9] have published guidance on security incident investigation and recovery techniques. The final phase of many security incident response approaches is the 'feedback' or 'follow-up' phase [7, 10, 11]. It is in this phase where an organization is expected to learn from a security incident with the aim of improving its overall security posture [10]. Within incident response, security incident learning typically accomplishes this through a series of formal reports, meetings and presentations to management after the closure of an incident investigation [7]. Lessons learned from a security investigation can include information about enhancements to existing security controls along with analyzing the necessity of changes to existing security incident response processes and procedures [12].

For an organization to learn more about the underlying causes of a security incident, investigators require access to detailed information [13-15]. In fact, ENISA contends that one of the critical factors influencing the success of an organization's security incident response team is the quality of actionable information at the disposal of the team [13]. However, researchers have observed that many organizations are more focused on eradication and recovery and less on security incident learning [16-18]. As a result, there is the potential that the quality of data derived from security investigations may be unfit for in-depth security incident learning. In practice, poor quality data produced by a security incident response process can also impact other aspects of cybersecurity within an organization [19].

In the past few years, researchers have argued that security incident response teams can provide much more functionality to an organization than just minimizing the damage from a security incident [19-21]. One particular function that has emerged is the role of a security incident response team within an organization's security threat intelligence program [19-21]. The primary objective of such a program is to produce evidence-based knowledge about risks and threats, which can then be used to make informed security decisions within an organization [22]. However, for such a security effort to be effective, it must be provided with datasets of sufficient quality [19]. Providing a threat intelligence program with either inaccurate, inconsistent or outdated information can produce poor quality intelligence [19].

From a threat intelligence perspective, security response teams provide organizations with an internal source of information. The data produced by these teams can include information like command and control IP addresses, low-level indicators of compromise, and malware hash values [13, 20]. Based on the increasing number of incidents, the importance of information from quality data sources, regulatory activity, and issues around eradication and recovery prompts the hypothesis that *organizations need to enhance the quality of data generated during security incident response investigations*. The proposed hypothesis raises the following research questions:

- What information is produced by a real-world security incident response team?
- Does a security incident response team face challenges when attempting to collect and document data during a security incident response investigation? If so, can these challenges be identified?

The contribution of this paper is a detailed case study examining the quality of data generated by a security incident response team within a Fortune 500 Financial organization. The case study encompasses a document analysis involving security investigation records, and interviews conducted with the organization's security incident response team. The remainder of this paper is structured as follows. Section 2 discusses relevant research related to security incident response and security threat intelligence. Section 3 describes the research methodology and introduces the case study. Section 4 describes the case study findings. Section 5 concludes the work conducted and presents ideas for future research.

## 2. Related Work

A growing number of researchers argue that security threat intelligence programs are becoming a fundamental component of an organization's broader security agenda [20, 23, 24]. McMillan summarizes security threat intelligence as evidence-based knowledge about threats, which can be used to make informed security decisions within an organization [22]. Brewer [25] adds that the objective of security threat intelligence is to deliver information, at the right time, with the correct and appropriate context, in order to reduce the amount of time it takes an organization to detect and respond to a security threat.

Mattern et al. [26] argue that organizations need to consider multiple sources of information for its security threat intelligence program. These sources can include data generated outside of an organization such as governmental projects, open source, and publicly-available databases, as well as commercial providers [21]. However, data for a security threat intelligence program can also be generated internally within an organization [10, 12, 21]. For example, network monitors, host-based indicators, and an organization's security incident response team [10, 12, 21]. The purpose of this team is to minimize the effects of an incident and manage an organization's return to an acceptable security posture [14]. However, this team can also contribute to an organization's security threat intelligence program by conducting detailed investigations, identifying root-causes associated with security events and incidents and producing actionable information [13]. This actionable information can include rogue IP addresses, malware metadata, and indicators of compromise [13, 20]. This information can also be of interest to regional and national Computer Emergency Response Teams [21, 27].

Regardless of the source, for data to be useful in a security threat intelligence program, it must be timely, actionable and relevant so that it can assist decisionmakers [26]. Previous research has focused on the quality of data in various security threat intelligence platforms [19, 28, 29]. Sillaber et al. [19] conducted a study that involved interviewing stakeholders responsible for security operations within large organizations. The purpose of the study is to investigate data quality challenges in threat intelligence sharing platforms. One of the findings from their research is that manually-entered information is very susceptible to data quality issues due to limited data entry checks. Sillaber et al. [19] go on to suggest that organizations implement automated data quality error checks for both internal and external threat intelligence sources. In concert with these findings, Al-Ibrahim et al. [29] proposed and evaluated various data quality dimensions for security threat intelligence platforms. These dimensions included correctness, relevance, utility, and uniqueness, which were then evaluated through an empirical case study using real-world data from antivirus scans [29]. Separately, Dandurand and Serrano [28] proposed eleven requirements concerning the sharing of threat intelligence. However, none of these requirements concerned the quality of data.

In response to a growing security threat, organizations are examining different security incident response approaches. Typically, these approaches consist of six phases [11, 12, 30]: *preparation*, which leads to the *detection* of an incident, followed by its *containment* which, in turn, allows security incident response teams to *eradicate*, *recover* and then, potentially, provide *feedback* information into the preparation phase. Although the literature has focused on the technical practices for implementing these phases within organizations, researchers have also identified that many organizations appear to find it difficult to apply these approaches [18, 31-33]. The implementation difficulty is apparent in various case studies undertaken within organizations.

Hove et al. [31] studied three large organizations with the purpose of examining the security incident handling plans and procedures within the studied organizations. Hove et al. [31] identified that based on best practices, many of the organizations were missing procedures. For example, in two of the organizations, security incident reporting procedures were not established while the other organization did not appear to have enough staff to respond to incidents efficiently [31]. Grispos et al. [34] studied how employees within an organization identify and report security incidents, based on the process that exists within the organization. The results of this study indicated that there are opportunities to improve security incident recognition and report generation within the organization, including education initiatives on 'what to do' and 'when to do it' in regard to incidents [34].

Grimes [32] argued that most security incident response models have become outdated and no longer support organization's efforts to respond to security incidents. Werlinger et al. [33] support these arguments and add that security incident handlers often need to develop their own tools to perform specific tasks during investigations. Tan et al. [18] focused on factors which influenced when an organization conducts an actual investigation, once a security problem is detected. As part of their case study, Tan et al. [18] reported that their studied organization had no precise definition for the term 'security incident' and as a result, incident handlers did not realize what security problems were actually 'incidents'. Tan et al. [18] found that this problem decreased the overall response time to an actual incident. While previous research has examined data quality challenges in security threat intelligence sharing platforms and the challenges of implementing security incident response processes within organizations, minimal research investigates the quality of data generated during a security incident response investigation.

## 3. Research Method

To empirically evaluate the quality of data generated during and after security incident response investigations, an exploratory case study was undertaken within a Fortune 500 Financial organization [35]. The benefit of conducting exploratory case studies is that they assist researchers to understand problems in real-world contexts, along with identifying future areas of research [35]. The case study was conducted between May and August 2013. The name of the organization is being withheld to ensure organizational anonymity. Therefore, the names of corporate documents and processes have been altered, and the results of data collected in the organization are presented anonymously. Maintaining organizational anonymity helps attain sensitive material, while creating an environment that is conducive to the presentation of this information.

The case study utilizes a mixed method approach to the collection of data [35]. During the case study, data was collected through an analysis of relevant security incident response documentation, internal security incident response investigation records, and through interviews with individuals within the organization. This data collection was undertaken in three phases.

The first phase involved analyzing relevant security incident response documentation. The analysis was performed to determine how management expects security incident response investigations to be conducted within the organization, as well as what information should be recorded during investigations. The primary author was given access to the organization's internal documentation repository, which was examined to identify and analyze documents related to security incident response processes, within the organization. These documents are available to all individuals within the organization's Information Security Incident Response (ISIR) team. Materials, which were considered sensitive, confidential and only available to management, were outside the scope of the analysis. The documents used in the exploratory case study were all signed-off by management before being stored in the document repository. The documents were analyzed using theme analysis, which allows a researcher to examine a variety of topics within a set of documents [35, 36]. In this case, the topics examined are relevant to security incident response settings.

The second phase of data collection involved the organization's security incident response database. The purpose of this analysis was twofold. First, it was used to examine if security investigations are managed and handled as per documented processes within the organization. Second, the database analysis was used to investigate the quality of data stored in the organization's security investigation records. For this analysis, the investigation records were examined from the perspective of four data quality dimensions: *accuracy*, *timeliness*, *completeness,* and *consistency* [37]. For the purpose of this research, accuracy is concerned with the difference between the correct information required to be documented in the investigation record and the information actually documented by the incident handlers. Timeliness refers to information that is in error because it is outdated and differs from the original value. Completeness is concerned with ensuring that no information is missing, while consistency implies that some form of standard exists throughout the information values. These dimensions have previously been used to examine the quality of data within information systems [37]. All the investigation records stored in the database were made available by the organization and are included in the analysis.

The third phase of data collection utilized semi-structured interviews with practitioners in the organization. The purpose of the interviews was to identify and explore challenges related to conducting security incident response investigations in the organization, along with examining information gathering throughout the overall response process. The interview instrument consisted of both open-ended and closed questions, which were derived from themes identified within industrial white-papers and academic research related to security incident response. To mitigate researcher bias regarding reliability and viability, the interview instrument was validated by two information security professionals [35]. Validation was only conducted once due to time constraints. The interview instrument was approved by the University of Glasgow Ethics Committee.

Initially, interviews were conducted with three individuals identified through the organization's security incident response process as the 'primary incident handlers'. A further twelve individuals were then identified and interviewed based on answers from the initial respondents' interviews. All fifteen individuals have at some point, been involved in the investigation and handling of a security incident within the organization. All responses to individual questions were initially recorded by hand and then digitally recorded soon after the interview completion, typically within an hour. The results were then examined by hand to identify trends, patterns, and anomalies.

The scope of this research is restricted from the following perspectives. This research consists of a single case study in a Fortune 500 Financial organization based in the United Kingdom. Hence, factors potentially impacting the case study include local, national and international regulatory requirements, along with societal and organizational cultural issues. It should also be noted that the primary researcher was embedded in the organization for several months as-well-as being the primary data collector for this investigation.

### 3.1. Information Security Incident Response (ISIR) Team

The organization's Information Security Incident Response (ISIR) team is an ad-hoc team of security incident handlers. This team facilitates the identification and assignment of actions required to prevent the recurrence of issues that are deemed to be or contribute to a security incident. The ISIR team follows a customized security incident handling approach. The approach comprises four phases: 1) incident detection and reporting; 2) recording, classification, and assignment; 3) investigation and resolution; and 4) incident closure.

The incident detection and reporting phase is concerned with the reporting of a security incident to the ISIR team. The reporting of an incident can come from one of the following sources: a direct request from senior management within the organization; a request from a member or management of the Information Security unit; a request from the Legal Services unit; or a request from the Human Resources department within the organization. The ISIR team can also be alerted about potential incidents by automated methods such as intrusion detection systems and data loss prevention systems.

During the recording, classification and assignment phase, the ISIR team will determine if a security incident really exists. If the incident is security-related, an investigation record is created in the ISIR team database. An incident handler is then appointed, and the team agree on the problem statement and the incident classification. Depending on the type and impact of the security incident, different stakeholders could be involved in the subsequent management and investigation. For example, if the security incident is determined to have a regulatory impact, a governance process is invoked together with the organization's risk unit, although the ISIR team still manages the incident.

The investigation and resolution phase of the process identifies the evidence and information that is required to conduct a security investigation. At this point, the ISIR team holds an incident meeting where the root-cause is established, and remedy actions associated with the incident are assigned to individuals. These individuals are expected to fulfill their actions and update the incident handler upon their completion.

The final phase, incident closure, involves two stages. First, relevant stakeholders are notified that all assigned actions have been completed and the security incident record is updated to reflect the closure of the incident. The second stage requires that the incident handler stores any findings and lessons learned acquired from the incident in the ISIR team database. At this point, the security incident is considered closed.

### 3.2. ISIR Team Database

Security investigation records are stored in the Information Security Incident Response (ISIR) team database, which is hosted on an IBM Lotus Notes server within the organization. Within this database, individual security investigation records are stored as separate documents. Each document includes a copy of the security investigation record template, as shown in Figure 1 – ISIR Record Template.

**Figure 1: ISIR Record Template**

The investigation record template consists of twenty-two (22) fields, divided into three parts (labeled as Parts A through C in Figure 1). These labels have been added to the record template to aid with the discussion below. Part A of the template prompts incident handlers to record information concerning the reported date and time of the security incident. The third field in this section called 'Duration', is used to document how long a particular security investigation took to complete within the organization.

Part B of the template concerns contact details about the individual who is managing and handling the investigation within the ISIR team. Recorded information includes the incident handler's name and job title; the name of their department and its physical location; their telephone and mobile phone numbers; email address and fax number.

Part C provides fields where incident handlers are expected to document information about the investigation itself. Although no confirmation is provided in the record template or within the ISIR process, the 'Date' and 'Time' fields within this section appear to be used to document the initial start date and time of the investigation. The purpose of the 'Incident Type' and 'Incident Location' fields is to document the type of investigation and its location in the organization. The 'Initial Impact assessment' and 'Incident Cause' fields are used to document any initial assessment of how the incident has affected the organization and what caused the incident to occur. However, the ISIR process does not elaborate on what information should be documented in these fields. The 'Investigation Record' field provides a space for the ISIR team to document and record investigation proceedings as and when they occur.

At the conclusion of an investigation, the incident handlers can complete the remaining fields at the bottom of Part C. The 'Cost of Incident' field can be used to record the resources expended on an investigation, while the 'Conclusion' field provides a space for the incident handler to document concluding remarks from the investigation. The final two fields 'Post Incident Lessons Learned' and 'Preventive Actions to be Taken' are used to document and record any lessons learned identified from the investigation, as well as any actions, which need to be taken post-incident.

### 4. Data Collection and Analysis

Fifteen (15) semi-structured interviews were conducted within the organization during the exploratory case study. The interview sample consisted of individuals in a variety of information security roles including, information security managers, senior security analysts, and security analysts. The individuals also indicated that they had a diverse range of work experience within a technical role. The interviewees have a minimum of two (2) years and a maximum of thirty-nine (39) years experience. The mean average experience of the interviewees was thirteen and a half (13.5) years.

The analysis of the ISIR team database revealed that one hundred and eighty-eight (188) security investigations were recorded in the database, at the time of the case study. The analysis of the investigation records, coupled with the findings from the interviews, are presented from the perspectives of data accuracy within the records and the timeliness of information available to the ISIR team. The records are also examined from the standpoint of documented information consistency and completeness.

### 4.1. Accuracy of Data

Regarding data accuracy, several observations appeared during the investigation records analysis. The first observation is related to the accuracy of date information within the investigation record itself. The 'Date' field is initially available in Section A of the record template, where the assumption is that the field describes the reported date of a security incident to the ISIR team. However, a 'Date' field is also present in Section C of the record template, with no further guidance on the purpose of this 'Date' information field. There are several potential uses of this field including the opening date for the investigation record, the incident identification or discovery date, or the implementation date of the first mitigation actions. As a result of this ambiguity, there is the potential that the information recorded in this 'Date' field is inaccurate. The information documented in the field might not represent the correct value that management expects within the investigation record template. The problem is inflated further due to a lack of document process specificity regarding what date information should be documented into the template.

A second observation related to the accuracy of data is related to the 'Incident Type' field. The analysis of the records revealed that the word 'incident' is prevalent in the majority of the investigation records. This strongly suggests that these records are considered by the ISIR team to be 'security incidents'. However, when queried about this phenomenon during the interviews, the participants indicated that a large number of the records classified as 'security incidents' were not all 'incidents' but a combination of 'security incidents' and 'security events'. It was interesting to observe that at the time of the case study, no formal security incident response categorization taxonomy existed within the organization. Hence, there is the potential that the number and type of incidents recorded as occurring within the organization is imprecise information.

The results, of querying the interview participants as to the meaning of the term 'security incident', supports the idea that the classification of security incidents is imprecise. A wide variety of answers were received, which included: "breach of security policy"; "degradation or circumvention of security controls"; "data loss"; "financial losses" and "threat to service availability". The variety of answers received from this query indicates that the organization does not have a unified definition of the term 'security incident'. The consequence is that the ISIR team could find it difficult to identify, generate and document accurate incident category titles for security investigations occurring within the organization.

A third observation concerning the accuracy of information in the investigation record regards the documentation of contact information. According to the investigation record template, the ISIR team is expected to use the fields within Section B of the template to record contact information regarding the incident handler assigned to the investigation. However, the analysis of this information revealed that the contact information for nearly half the records did not belong to individuals handling the incident, but the individuals who are reporting the incident to the ISIR team. It is imperative to note that the documented process does not specify what information should be recorded in the template. However, this ambiguity has resulted in the storage of inaccurate information in Section B of the investigation record template. Documentation clarity regarding the type of information to record in this section would help the ISIR team to quickly identify whom to contact for further details regarding specific investigations. Hence, more precious and accurate information could be recorded in the investigation record.

### 4.2. Timeliness of Data

Time is a critical factor in many security incident investigations [11, 16]. In some incidents, it could become vital for security incident handlers to quickly obtain access to data before it is overwritten or becomes outdated. However, during the interviews, ten (10) out of the fifteen (15) respondents suggested that the ISIR team often encounters challenges related to the timeliness of security data. As a result, some of this data could be outdated, or it may differ from the original value. A variety of challenges were described preventing access to timely data including obstacles to obtaining physical access to data, short data retention times, and limited support from third-parties involved in the incident investigation.

West-Brown et al. [38] argue that metrics provide an accurate way of quantifying the performance of security incident response teams. West-Brown et al. [38] defined response time as the period from the first report of a security incident to the implementation of the first mitigating actions. They describe the total time to resolve an incident as the time from when the security incident is reported to the time the incident is closed. It is interesting that West-Brown et al. [38] note that although such information is useful to analyze the historical performance of a security incident response team, there are no published recommended times to evaluate such groups. However, the information recorded in the investigation records can be useful to determine and identify potential challenges to obtaining timely information during security incident investigations.

Fifty-two (52) incident records contained data concerning the response time. The minimum response time calculated from these records was two minutes and the maximum response time was 325 minutes. The average response time was 56.30 minutes. Figure 2 – Cumulative Response Times presents the percentage of incidents which were responded to over a given period of time. As shown in Figure 2, the ISIR team takes mitigating actions within thirty minutes for approximately sixty percent (60%) of the recorded security incidents and within two hours for ninety percent (90%) of the incidents.

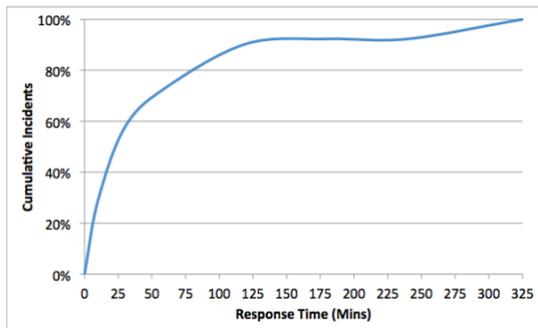

**Figure 2: Cumulative Response Times**

Sixty-two (62) of the incident records contain data concerning the total time to resolution; meaning that there were one-hundred and twenty-six (126) incidents that did not include enough information for this calculation. Out of the sixty-two (62), the minimum time to resolve an incident was half a day and the maximum time to address an incident was one-hundred and thirty (130) days. Therefore, the average time to address an incident was just under 12 days. Figure 3 – Cumulative Total Time to Resolve Incidents presents the percentage of incidents that were resolved from the first reporting to closure over a given period.

The ISIR team resolves twenty percent (20%) of the analyzed incidents in half a day, and eighty percent (80%) of the analyzed incidents are resolved within twenty days (20). In summary, the response and total time to resolve calculations support the idea that some of the investigation data being collected by the ISIR team could be either outdated or unavailable because of challenges preventing access to timely data.

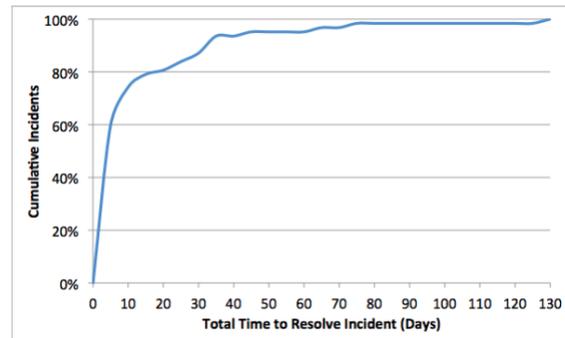

**Figure 3: Cumulative Total Time to Resolve Incidents**

### 4.3. Completeness of Data

The completeness analysis focuses on the number of fields within the investigation record that are completed during the documentation of a security investigation. At a high-level, this analysis shows that only one field, the 'Investigation Record' field, was considered complete in the one hundred and eighty-eight (188) investigation records. For this analysis, a field is considered complete if data is visible in the field, regardless of its accuracy dimension.

A further seven (7) out of the twenty-two (22) fields were completed in at least ninety-four percent (94%) of the investigation records. In contrast, the 'Duration' and 'Cost of Incident' fields were only completed in thirteen percent (13%) of the investigation records. Moreover, fifty (50) investigation records (27%) contained information within the 'Post-Incident Lessons Learned' field, and only twenty-eight (28) records or fifteen percent (15%) provide information within the 'Preventive Actions' field. Hence, nearly three-quarters of the investigation records did not contain lessons learned regarding the security investigation. Table 1 – Number of Completed Fields summarizes the number and percentage of completed fields within the ISIR team database.

From an overall investigation record perspective, only one (1) out of the one hundred and eighty-eight (188) records analyzed during the study were considered completed, from the perspective of all twenty-two (22) fields. This means that one hundred and eighty-seven (187) records contained one or more fields that were incomplete, regarding information in the documented security investigation. Fifteen (15) investigation records were found to be missing between one and three fields; forty-one (41) records were found to be missing four to six fields, while one hundred and seventeen (117) records were missing seven to ten fields of information. Furthermore, fourteen (14) records were missing information within eleven (11) or more fields in the investigation record template. This finding suggests that the incident handlers are not completing the entire incident record during incident investigations within the organization.

| Field Name | No. of Records (% of Overall) |
|---|---|
| Date | 181 (96%) |
| Time | 176 (94%) |
| Duration | 24 (13%) |
| Name | 186 (99%) |
| Job Title | 185 (98%) |
| Department | 184 (98%) |
| Location | 180 (96%) |
| Telephone | 180 (96%) |
| Mobile | 62 (33%) |
| Email | 164 (87%) |
| Fax | 39 (21%) |
| Date | 178 (95%) |
| Time | 167 (89%) |
| Incident Type | 145 (77%) |
| Incident Location | 97 (52%) |
| Initial Impact Assessment | 68 (36%) |
| Incident Cause | 155 (82%) |
| Investigation Record | 188 (100%) |
| Cost of Incident | 25 (13%) |
| Conclusion | 54 (29%) |
| Post-Incident Lessons Learned | 50 (27%) |
| Preventative Actions to Be Taken | 28 (15%) |

**Table 1: Number of Completed Fields**

To explore why incomplete information was recorded in the investigation records, the interview participants were queried to determine if and when data is collected regarding a security investigation. The majority of the answers returned were positive. Fourteen (14) out of the fifteen (15) respondents suggested that data is collected and stored throughout the organization's security incident response lifecycle. However, one individual indicated that he/she did not know if the practice took place. The respondents did indicate that investigation information was assigned and performed by an incident handler that has been designated as the primary incident handler and it was the responsibility of this individual to ensure that information was documented.

When the individuals were asked about what information is documented within the records, this information included investigation meeting notes, actions to be taken for remediation, copies of any logs and emails associated with the investigation, as well as communication between the ISIR team and management. However, information related to the individual fields within the investigation record template were not mentioned by the respondents. This suggests that the organization does not have a uniform approach to capturing specific information and that the information captured focuses on assisting with the eradication and recovery aspect of the lifecycle. However, this approach potentially hinders capturing actionable information that could facilitate meaningful threat intelligence for future use.

### 4.4. Consistency of Data

Regarding consistency, the analysis focused on the information recorded within the individual fields in the investigation record template. Several observations were evident in this analysis. The first observation was related to the representation of date and time information within the investigation records. In both cases, the document analysis revealed that incident handlers are not provided guidance on a standard format to document these two metrics. As a result, dates and times are stored in different formats. For example, date information was found to be in the following formats: DD/MM/YY, DD/MM/YYYY, DD/MM, and MM/DD. Similar observations were noticed with time information, which was documented in both 12-hour and 24-hour formats. Moreover, in some cases, it was difficult to distinguish what time format was actually being used by the incident handler.

The second observation regarding consistency is related to the 'Incident Type' field, and the inconsistent information used to describe an incident. For example, an investigation where potential data loss has been an issue, the following information was found to be recorded in the 'Incident Type' field': 'potential data exposure', 'potential data leakage', 'potential security breach', 'exposure of live data', 'email to the wrong person', and 'loss of data'. This finding reiterates the lack of a consistent security incident response categorization taxonomy within the organization. However, it also revealed that two investigation records of the same type, can potentially, be described with different incident type descriptions. This could complicate the identification of indicators of compromise later in the intelligence process.

One potential explanation regarding the inconsistency of data recorded in the investigation record could be the limited resources available to the ISIR team. While twelve (12) out of the fifteen (15) individuals confirmed that the organization uses a document-centric security incident response approach, five (5) people suggested that the process is not always followed. When questioned as to the reasons for process deviation, answers comprised of time constraints, a lack of staff to run the entire process, and a lack of support for handling specific security investigations. Hence, the findings suggest that there is a potential link between the quality of consistent information documented in the incident investigation record and the resources available to the ISIR team.

## 5. Conclusions and Future Work

Security incidents are increasingly impacting industries in today's globally networked environments. As a result, organizations are exploring alternative security measures, such as security threat intelligence programs. However, for such a program to be a success, organizations need to ensure that all viable data is collected, validated, and recorded into relevant datastores, to support an overall security threat intelligence effort. Security incident response investigations are no exception to these requirements. Any data collected during the course of an investigation that is translated into actionable information will have a direct impact on the viability of derived intelligence. Empirical analysis of the security incident response landscape in a Fortune 500 Financial organization revealed that the quality of data generated during the security incident response lifecycle needs to be addressed. The results from the case study suggest that the data currently generated from the organization's security incident response process does not appear to help facilitate security threat intelligence, either during or after the closure of an investigation. A message that emerges from this research is that other organizations, similar to the organization in this exploratory case study, need to examine the quality of data generated during and after their security incident investigations. It is plausible that other organizations could also have data quality issues in their security incident response lifecycle.

Several opportunities exist for future research. One area of future research will identify, investigate and evaluate methods, tools, and techniques that can be used to enhance the quality of data generated during and after security investigations. One potential avenue could involve the integration of disciplined agile principles and practices into the security incident response process as a viable method of strengthening an organization's security incident response data quality proficiencies. Another potential area of research is to explore automated approaches for reducing data quality issues during a security incident response investigation. Additional areas of research also include investigating the optimal data capture automation and the overall impact of this automation throughout all the phases of a security incident investigation. This research would include the examination of methods and tools to improve the efficiency and effectiveness of data generation and collection within security incident response teams. Future research will repeat and expand the existing case study in similar and other organizations and industry sectors. The objective of the expansion is to determine the generalizability and transferability of the data quality issues identified in this case study with respect to other industries.